\documentclass[sigconf]{acmart}

\AtBeginDocument{%
  \providecommand\BibTeX{{%
    \normalfont B\kern-0.5em{\scshape i\kern-0.25em b}\kern-0.8em\TeX}}}

\usepackage{amsfonts}
\usepackage{algorithm}
\usepackage{algcompatible}

\usepackage{graphicx} 
\usepackage{float}    
\usepackage{subfigure}
\setcopyright{none}
\renewcommand\footnotetextcopyrightpermission[1]{}
\usepackage{enumitem}
\usepackage{amsmath}
\usepackage{booktabs}
\usepackage{multirow}
\usepackage{multicol}
\usepackage[normalem]{ulem}
\useunder{\uline}{\ul}{}
\usepackage{appendix}
\usepackage{balance}
\usepackage{makecell}
\usepackage{color}
\usepackage[marginal]{footmisc}

\definecolor{bgcolor}{RGB}{242, 242, 242}
\usepackage{caption}
\usepackage{marvosym}

\newsavebox{\promptboxbox}
\newenvironment{promptbox}
 {\begin{lrbox}{\promptboxbox}\begin{minipage}{\dimexpr\columnwidth-2\fboxsep}\itshape}
 {\end{minipage}\end{lrbox}%
  \par\medskip\noindent
  \colorbox{bgcolor}{\usebox{\promptboxbox}}%
  \par\medskip}

\copyrightyear{2025}
\acmYear{2025}
\setcopyright{acmlicensed}
\acmConference[CIKM '25] {Proceedings of the 34th ACM International Conference on Information and Knowledge Management}{ November 10--14, 2025}{Seoul, Republic of Korea.}
\acmBooktitle{Proceedings of the 34th ACM International Conference on Information and Knowledge Management (CIKM '25), November 10--14, 2025, Seoul, Republic of Korea}
\acmISBN{979-8-4007-2040-6/2025/11}
\acmDOI{10.1145/XXXXXX.XXXXXX}

\begin{document}

\title{Harnessing Large Language Models for Group POI Recommendations}

\author{Jing Long}
\email{jing.long@uq.edu.au}
\orcid{1234-5678-9012}
\affiliation{%
 \institution{The University of Queensland}
 \city{Brisbane}
 \postcode{4072}
 \country{Australia}}

\author{Liang Qu}
\email{liang.qu@uq.edu.au}
\orcid{1234-5678-9012}
\affiliation{%
 \institution{The University of Queensland}
 \city{Brisbane}
 \postcode{4072}
 \country{Australia}}

\author{Junliang Yu}
\email{jl.yu@uq.edu.au}
\affiliation{%
 \institution{The University of Queensland}
 \city{Brisbane}
 \postcode{4072}
 \country{Australia}}

 \author{Tong Chen}
\email{tong.chen@uq.edu.au}
\affiliation{%
 \institution{The University of Queensland}
 \city{Brisbane}
 \postcode{4072}
 \country{Australia}}

\author{Quoc Viet Hung Nguyen}
\email{henry.nguyen@griffith.edu.au}
\affiliation{%
 \institution{Griffith University}
 \city{Gold Coast}
 \state{QLD}
 \postcode{4222}
 \country{Australia}}
 
\author{Hongzhi Yin*}
\email{h.yin1@uq.edu.au}
\affiliation{%
 \thanks{*Corresponding author}
 \institution{The University of Queensland}
 \city{Brisbane}
 \postcode{4072}
 \country{Australia}}

\renewcommand{\shortauthors}{Jing Long et al.}

\begin{abstract}
The rapid proliferation of Location-Based Social Networks (LBSNs) has underscored the importance of Point-of-Interest (POI) recommendation systems in enhancing user experiences. While individual POI recommendation methods leverage users' check-in histories to provide personalized suggestions, they struggle to address scenarios requiring group decision-making. Group POI recommendation systems aim to satisfy the collective preferences of multiple users, but existing approaches face two major challenges: diverse group preferences and extreme data sparsity in group check-in data. To overcome these challenges, we propose LLMGPR, a novel framework that leverages large language models (LLMs) for group POI recommendations. LLMGPR introduces semantic-enhanced POI tokens and incorporates rich contextual information to model the diverse and complex dynamics of group decision-making. To further enhance its capabilities, we developed a sequencing adapter using Quantized Low-Rank Adaptation (QLoRA), which aligns LLMs with group POI recommendation tasks. To address the issue of sparse group check-in data, LLMGPR employs an aggregation adapter that integrates individual representations into meaningful group representations. Additionally, a self-supervised learning (SSL) task is designed to predict the purposes of check-in sequences (e.g., business trips and family vacations), thereby enriching group representations with deeper semantic insights. Extensive experiments demonstrate the effectiveness of LLMGPR, showcasing its ability to significantly enhance the accuracy and robustness of group POI recommendations.

\end{abstract}

\maketitle

\section{Introduction}
With the rapid expansion of Location-Based Social Networks (LBSNs) such as Foursquare and Weeplace, Point-of-Interest (POI) recommendations have become essential for helping individuals discover places that align with their interests \cite{zhang2025survey,yin2015joint,huang2024counterfactual}. These recommendations are crucial for applications like mobility prediction, route planning, and location-based advertising \cite{Li2018NextPR, yin2016spatio,yin2016joint}. While individual POI recommendation systems effectively cater to personal preferences by utilizing users' check-in histories, they fall short in scenarios requiring group decision-making. As location-based social activities become more prevalent, the demand for systems that address the preferences of multiple users simultaneously has transformed into a pressing need. This has given rise to the concept of Group POI Recommendation \cite{li2020group,liu2024poi}, which is increasingly essential in scenarios such as friends planning a weekend outing, families on vacation, or colleagues organizing a business trip, where satisfying the collective preferences of all group members is crucial.


\begin{figure}
    \centering
	\includegraphics[width=\linewidth]{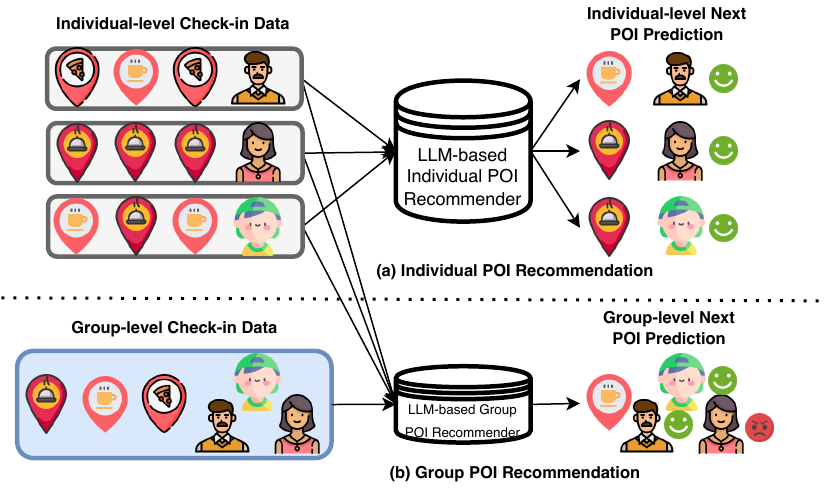}
	\caption{(a) Example of individual POI recommendation. (b) Example of group POI recommendation.}
	\label{Fig:toyexample} 
\end{figure}

Group POI recommendation systems typically involve a two-phase pipeline. In the first phase, the group is treated as a virtual user, allowing traditional individual POI recommendation methods, such as matrix factorization \cite{2014GeoMF}, transformer-based approaches \cite{2021STAN}, graph-based approaches \cite{10.1145/3477495.3532012}, and diffusion-based approaches \cite{Qin_2023}, to be directly applied. These methods learn group representations based on group-level check-in data and individual representations based on individual-level check-in data, respectively. 
In the second phase, these group representations are further enhanced by aggregating the individuals' representations within the group. Aggregation approaches include the simple average \cite{vinh2019interact}, weighted average \cite{sojahrood2021poi}, attention mechanism \cite{sankar2020groupim}, and graph convolution network \cite{li2024multi}. However, despite their effectiveness, existing group POI recommendation systems predominantly rely on ID-based data, utilizing only sparse individual and group check-in data to capture statistical patterns, leaving the semantic richness inherent in POI text data untapped.

Benefiting from the recent breakthroughs in pre-trained large language models (LLMs), which have significantly advanced the semantic understanding of natural language, researchers have begun exploring the integration of LLMs into POI recommendation systems. This integration aims to better capture the rich semantic information embedded in POI data (e.g., POI categories), thereby alleviating data sparsity and cold-start issues. For example, LLMMove \cite{feng2024move} and LLM-Mob \cite{wang2023would}  utilize a series of carefully designed prompts to input individual users' check-in records into pre-trained LLMs, leveraging the commonsense knowledge embedded in the LLMs to better exploit the rich semantic information in check-in data.  Additionally, LLM4POI \cite{li2024large} further fine-tunes the pre-trained LLM using individual check-in data to enhance its performance. 

Although existing LLM-based POI recommendation methods have shown effectiveness, they are primarily designed for individual POI recommendations and face significant challenges when applied to the group POI recommendation scenario. These challenges arise from two major issues: 
\begin{itemize}[leftmargin=*]
    \item \textbf{Diverse Preferences and Complex Group Decision-making}. Unlike individual POI recommendations, which aim to satisfy the preferences of a single user (as illustrated in Figure \ref{Fig:toyexample}), group POI recommendations require reaching a consensus among group members on a shared POI to visit. This is particularly challenging because the preferences of individuals within a group are often diverse, and, in some cases, the group’s overall preference may differ from those of its individual members \cite{guo2021hierarchical}. Modeling this intricate decision-making process is a complex task.
    \item \textbf{Extreme Data Sparsity}. Group check-in data is inherently much sparser than individual check-in data, as it involves multiple users visiting the same POI simultaneously (e.g., a group of friends dining together at a restaurant). For example, as discussed in Section \ref{sec:dataset}, group check-in data in three public datasets are approximately 50–100 times sparser than individual check-in data. While existing LLM-based POI recommendation methods have demonstrated potential in mitigating data sparsity at the individual level by utilizing semantic information, the extreme sparsity of group check-in data presents an even greater challenge. 
\end{itemize}

To address the challenges of group POI recommendation, we propose LLMGPR, a novel framework that leverages pre-trained large language models (LLMs) to empower Group POI recommendation. To tackle the diverse group preferences challenge, our framework first introduces semantic-enhanced POI tokens, which are integrated alongside the original word tokens of the LLM. The embeddings for these POI tokens are initialized by extracting rich textual information (e.g., category and address) using the LLM. By combining semantic-enhanced POI tokens with contextual information such as positional encodings, spatio-temporal differences, and sequential patterns, our model captures the multifaceted factors influencing group decision-making. These include sequential effects, geographical influences, temporal patterns, and semantic relationships \cite{xie2016learning}. Building on this, we propose a sequencing adapter based on Quantized Low-Rank Adaptation (QLoRA), which processes the representations of check-in activities to generate check-in sequence representations. These representations can then be flexibly applied as either group-level or individual-level representations, depending on the sequence type. 

To address the extreme data sparsity challenge in group check-in data, we introduce two key components. First, an aggregation adapter combines individual member representations derived from their respective check-in sequences into a unified group representation. Second, we design a self-supervised learning (SSL) task to further enrich the group representations. This task predicts the underlying purposes of check-in sequences (e.g., business trips and family vacations) by utilizing spatio-temporal and contextual cues within each sequence. Specifically, we employ a purpose-oriented prompt to guide the pre-trained LLM in categorizing sequences, which then serves as ground truth for self-supervised learning. The SSL signal enhances the model’s ability to extract semantic information from sparse group-level check-in data, refining group representations.

By jointly optimizing the SSL task and the adapters for the recommendation task, LLMGPR effectively leverages the rich semantic and contextual information in check-in data, achieving robust and accurate group POI recommendations.

The primary contributions of this work are summarized as follows:
\begin{itemize}[leftmargin=12pt]
\item  We propose LLMGPR, the first pre-trained LLM-based framework for group POI recommendation. 

\item We introduce semantic-enhanced POI tokens and incorporate rich contextual information to model group check-in activities. Additionally, we propose a sequencing adapter based on QLoRA to learn group representations from check-in sequences.

\item  We address the challenge of sparse group-level check-in data by developing an aggregation adapter to combine individual representations into group representations and introducing a SSL task that predicts check-in sequence purposes, enabling the model to extract meaningful semantic insights.

\item We conduct extensive experiments to demonstrate the effectiveness of the proposed LLMGPR, showing significant improvements in accuracy for group POI recommendations.
\end{itemize}

\section{PRELIMINARIES}\label{sec:prelim}

This section lists key notations used throughout this paper, outlines our primary task, and introduces Large Language Models and Quantized Low-Rank Adaptation as the background knowledge.

\subsection{Notations}
We denote the sets of users, POIs, categories, and groups as
$\mathcal{U}$, 
$\mathcal{P}$, 
$\mathcal{C}$, 
and $\mathcal{G}$
respectively. Each POI $p\in \mathcal{P}$ is associated with a category tag (e.g., entertainment or restaurant) $c_p\in \mathcal{C}$ and coordinates $(lon_p,lat_p)$. In addition, each group $g_n\in \mathcal{G}$ contains $K$ members, denoted as $\mathcal{K}_n=\{u_1,u_2,...,u_K\}$.

\textbf{Definition 1: Check-in Sequence}. A check-in activity indicates a user $u\in \mathcal{U}$ or a group $g\in \mathcal{G}$ has visited a POI $p\in \mathcal{P}$ at the timestamp $t$. By sorting all check-ins of the user or the group chronologically, a check-in sequence is obtained which contains $M$ consecutive POIs, denoted by $\mathcal{X}_u=\{p_1, p_2,...,p_{M}\}$ for the user $u$ or $\mathcal{X}_g=\{p_1, p_2,...,p_{M}\}$ for the group $g$.

\textbf{Definition 2: POI Token}. As the extensions of the original word tokens, POI tokens encapsulate unique IDs and encoded representations corresponding to specific POIs. To enrich these tokens with semantic information, their embeddings are initialized by analyzing related information with the large language model, and the details are shown in Section \ref{sec:lept}. All POI embeddings are stored in the POI embedding matrix $\textbf{E}_{poi}$.

\subsection{Task: Group POI Recommendation}

Given a group $g_n$, its check-in sequence $\mathcal{X}_{g_n}$, and individual check-in sequences of all group members $\mathcal{X}(\mathcal{K}_n) = \{\mathcal{X}_{u_1}, \mathcal{X}_{u_2},...,\mathcal{X}_{u_K}\}$, our goal is to generate a ranked list of possible POIs that group $g_n$ would like to visit next.

\subsection{Large Language Models}
Pre-trained large language models (LLMs) like GPT, T5, and Llama are advanced AI systems based on the Transformer architecture \cite{vaswani2017attention}, which uses a self-attention mechanism to capture long-range dependencies in text. This work is based on Llama3-8b \cite{touvron2023llama}, where input text sequences are tokenized, embedded, and processed through multiple Transformer blocks with layers for multi-head attention, feedforward transformations, and normalization. The model then generates tokens sequentially, producing coherent and contextually relevant text based on the given input and context.

\subsection{Quantized Low-Rank Adaptation}\label{sec:qlora}

Quantized Low-Rank Adaptation (QLoRA) \cite{dettmers2024qlora} enhances LLM efficiency by quantizing model weights and applying low-rank adaptations for task-specific fine-tuning. \textbf{Quantization} reduces weight precision, lowering memory and computational needs. For a linear weight matrix $\textbf{W}$, the quantized version $\textbf{W}_q$ is:
\begin{equation}
\textbf{W}_q = round(\frac{\textbf{W}-min(\textbf{W})}{\Delta}),
\end{equation}
where $\Delta$ is the quantization step size:
\begin{equation}
\Delta = round(\frac{max(\textbf{W})-min(\textbf{W})}{2^b - 1}),
\end{equation}
with $b$ being the quantization bits. \textbf{Low-Rank Adaptation} modifies $\textbf{W}_q \in \mathbb{R}^{d\times d}$ by:
\begin{equation}
\textbf{W}_q = \textbf{W}_q + \textbf{A} \cdot \textbf{B},
\end{equation}
where $\textbf{A} \in \mathbb{R}^{d\times r}$ and $\textbf{B} \in \mathbb{R}^{r\times d}$ are low-rank matrices updated during fine-tuning, and $r$ is the rank controlling adapter sizes. QLoRA is particularly useful for optimizing the LLM across tasks while conserving resources.

\begin{figure*}
	\includegraphics[width=\linewidth]{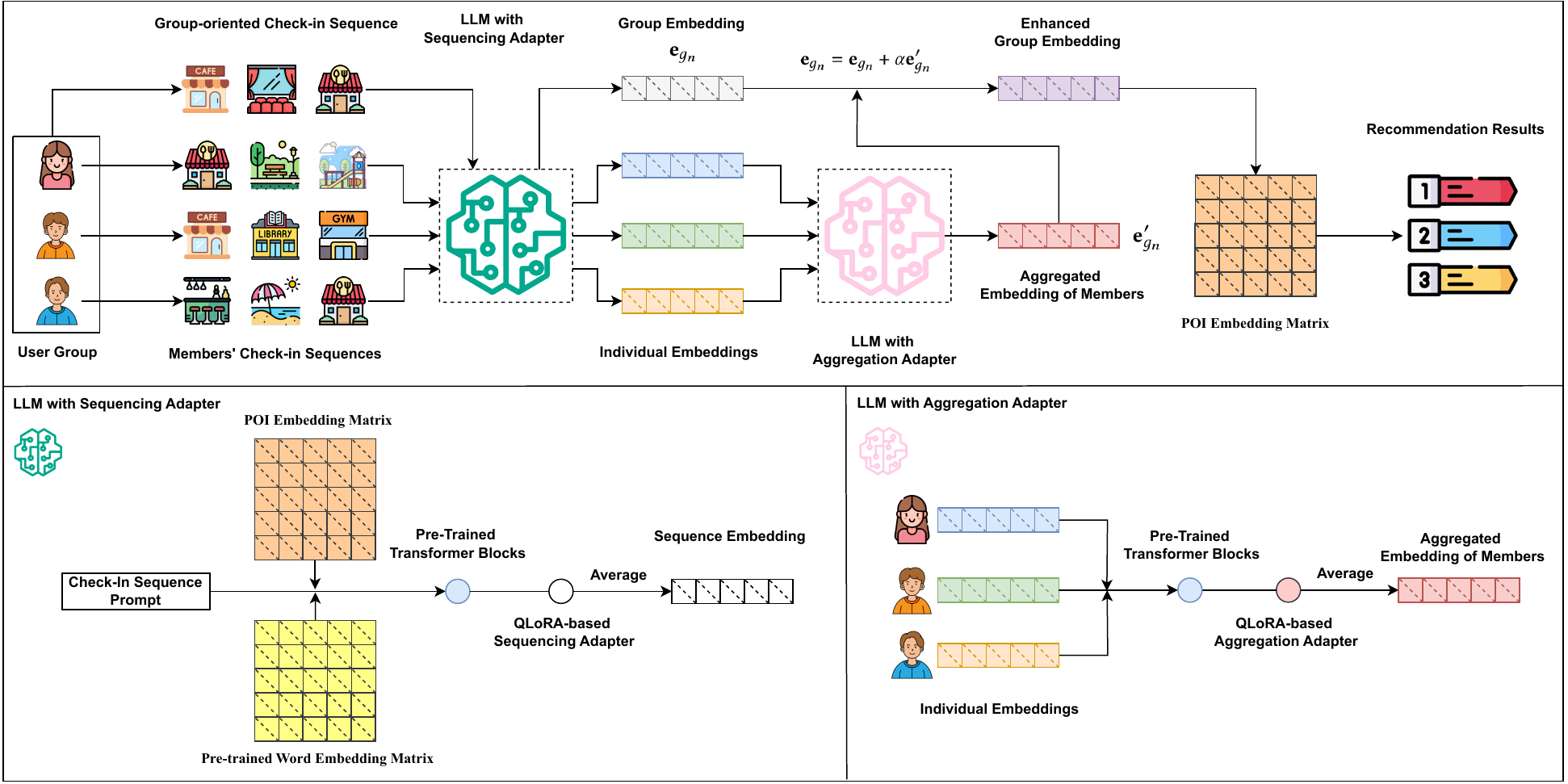}
	\caption{The overview of our proposed LLMGPR.}
    \vspace{-1.0em}
	\label{fig:overview}
\end{figure*}

\section{Methodology}
In this section, we formally introduce the design of LLMGPR, with an overview provided in Figure \ref{fig:overview}. The main components include: (1) \textbf{Learning group representation} under the guidance of the LLM. (2) \textbf{Enhancing group representation} by leveraging the LLM to aggregate preferences of group members. (3) A \textbf{self-supervised learning task} to further enhance group representations.

\subsection{Group POI Recommendation with LLM}\label{sec:lse}
The proposed framework introduces a methodology for learning group representations based on group check-in sequences, which is crucial for capturing group preferences and making high-quality recommendations. Existing methods for modeling check-in sequences have the following limitations. Firstly, they often overlook the semantic information embedded in POI names and categories, failing to capture the sequential effect of check-in activities at the semantic level. Secondly, they ignore the spatial and temporal differences between check-in activities, missing out on capturing precise contextual information. For example, when there is a large spatial or temporal difference between two consecutive check-ins, the earlier check-in should not be considered as context for the later one. This is a key difference from word sequences. Thirdly, these methods usually require extensive retraining to adapt to new groups or regions, limiting their scalability and practical application.

To address these limitations, we introduce the large language model, which can leverage its world knowledge to enhance the POIs' representations by exploiting their names and categories. By encoding temporal and spatial differences between check-in activities, the LLM captures the dynamic nature of user movements and preferences, resulting in richer and more context-aware representations. It excels at modeling sequential data, capturing long-range dependencies and contextual relationships between check-in activities. This allows for a more nuanced understanding of user behavior and improves prediction accuracy. Moreover, pre-trained on vast data from diverse domains, the LLM generalizes well to new users and regions without extensive retraining, making them more adaptable and scalable.

\subsubsection{Learning Embeddings of POI Tokens}\label{sec:lept}

Before establishing the sequence representation semantically, we first introduce POI tokens and initial corresponding embeddings alongside the original word tokens of the LLM. While semantic representations capture the nuanced meanings of POI names, categories and other information, relying solely on semantics may not fully leverage the unique identifiers of POIs. By incorporating POI embeddings through their IDs, we can ensure that each location is distinctly recognized and appropriately contextualized. This method allows us to directly reference specific POIs, enhancing the precision of recommendations and maintaining the integrity of location-specific data. To balance the depth of semantic understanding with the specificity of location identifiers, the embedding of each POI is initialized by the large language model with a carefully designed prompt which is shown as below:
\small
\setlength{\fboxsep}{5pt} 
\setlength{\fboxrule}{0pt} 
\noindent 
\begin{promptbox}
\textit{\textcolor{purple}{Generate an embedding for the following Point of Interest (POI).} \\
Name: \textcolor{teal}{<name>}\\
Category: \textcolor{teal}{<category>}\\\
Description: \textcolor{teal}{<description>}\\\
Reviews: \textcolor{teal}{<reviews>}\\\
Latitude: \textcolor{teal}{<latitude>}\\
Longitude: \textcolor{teal}{<longitude>}\\
Address: \textcolor{teal}{<address>}}
\end{promptbox}
\noindent{where} we directly use the average of the final-layer token embeddings as the initialized POI embedding without going through the linear and softmax layers. Applying this process to all POIs results in an embedding matrix $\textbf{E}_{poi}\in\mathbb{R}^{P \times d}$, where $P$ is the number of all POIs and $d$ is the embedding size which is the same as the size of the token embedding. In addition, these embeddings will be further fine-tuned during the following group POI recommendation task with specific objectives.

\subsubsection{Generating Candidate Group POI}
Given all POI embeddings, we design another prompt shown as follows to generate the representation of the check-in sequence. 
\small
\setlength{\fboxsep}{5pt} 
\setlength{\fboxrule}{0pt} 
\noindent 
\begin{promptbox}
\textit{\textcolor{purple}{Generate an embedding for the provided check-in sequence, which has multiple check-in
activities, and the format of a check-in activity is:} \\
(POIID, timestamp, temporal difference from previous check-in, spatial difference from previous
check-in).\\
The check-in sequence is: \textcolor{teal}{<check-in sequence>}}
\end{promptbox}
\noindent When this prompt is processed by the word embedding layer, all "POIIDs" of check-in activities will be regarded as special POI tokens instead of standard word tokens. Therefore, they are encoded by querying the POI embedding matrix $\textbf{E}_{poi}$ with their IDs. These POI embeddings, along with the embeddings of other standard word tokens, are then fed into the subsequent layers of the large language model (LLM) for further processing. Then, the average of the final-layer token embeddings will be used as the embedded representation of the check-in sequence, denoted as:
\begin{equation}
  \textbf{e}_{g_n} = AVG(\Theta(\mathcal{X}_{g_n},\textbf{E}_{poi})),
\end{equation}
where $\textbf{e}_{g_n} \in \mathbb{R}^d$ is the embedding of the group check-in sequence $\mathcal{X}_{g_n}$, and $\Theta$ refers to the original large language model. Given the group embedding $\textbf{e}_{g_n}$ and the embeddings of $h$ candidate POIs $\textbf{E}_{cand}=[\textbf{e}_{p_1},\textbf{e}_{p_2}
,...,\textbf{e}_{p_h}]\in\mathbb{R}^{h\times d}$, the likelihood $\textbf{a}\in\mathbb{R}^{h}$ of candidate POIs to be visited next by this group is calculated as:
\begin{equation}\label{eq:fr}
  \textbf{a} = Softmax(\textbf{E}_{cand}\textbf{e}_{g_n}^T).
\end{equation}

\subsubsection{Aligning LLM with Group POI Recommendation}
Fine-tuning a large language model for POI  recommendation, rather than relying on zero-shot learning, is essential because it tailors the model to the specific nuances and requirements of the task. This adaptation enhances relevance and accuracy by allowing the model to better understand domain-specific data and tasks. By integrating the specialized check-in sequence data, fine-tuning significantly improves the model’s performance, resulting in more accurate recommendations. Hence, we further employ the Quantized Low-Rank Adaptation (QLoRA)  to fine-tune the LLM for the specific task of generating check-in sequence representations. Specifically, QLoRA is applied to all linear layers in $\Theta$, while the detailed approach is described in Section \ref{sec:qlora}. 

Here, we use $\Theta^{'}_s$ to denote the adapter containing all additional weights from QLoRA for generating check-in sequence representations. During the fine-tuning process, the original LLM $\Theta$ is frozen while the sequencing adapter $\Theta^{'}_s$ and the POI embedding matrix $\textbf{E}_{POI}$ are updated by the POI prediction objective:
\begin{equation}\label{eq:lpoi}
  \mathcal{L}_{poi} =\mathcal{L}_{CE}(y',y),
\end{equation}
where ${L}_{CE}$ is the cross-entropy loss, $y'$ is the predicted POI based on Equation \ref{eq:fr}, and $y$ is the ground truth POI. For a check-in sequence $\mathcal{X}$, the predictions are made successively on sequences $\{p_1\}, \{p_1, p_2\},..., \{p_1,$ $p_2,...,p_{M_i-1}\}$, and $\{p_2, p_3, ..., p_{M_i}\}$ is the set of corresponding ground truth POIs. 
Due to the limited number of group check-in sequences, both group and individual check-in sequences are utilized to optimize the sequencing adapter and POI embeddings.

\subsection{Enhancing Group Representations with Individual Check-ins}\label{sec:er}
Even with the modified LLM, it is infeasible to accurately learn the group representation from limited group check-in data. It is essential to integrate and aggregate each individual's preferences (i.e., representation) within a group to enhance the group representation.  Existing methods like averaging individual representations \cite{vinh2019interact} or assigning weights based on importance or preference consistency \cite{yin2020overcoming} offer varying degrees of flexibility but fail to model the complex interaction patterns among group members. 

\subsubsection{Aggregating Individual Rereprentations with LLM}
In light of this, we employ the LLM to aggregate individual representations, leveraging its ability to capture intricate relationships and interactions, leading to a more nuanced and accurate group representation. Specifically, for the group $g_n$, we first obtain the group check-in sequence representation $\textbf{e}_{g_n}$ by applying the modified LLM $\Theta_s$ to the group check-in sequence $\mathcal{X}_{g_n}$ as described in Section \ref{sec:lse}. Then, for each member within the group $u_k \in \mathcal{K}_n$, we generate their individual representation from their check-in sequence $\mathcal{X}_{u_k}$:
\begin{equation}
  \textbf{e}_{u_k} = \Theta_s(\mathcal{X}_{u_k}).
\end{equation}
Hence, the representations of all individual check-in sequences are denoted as $\textbf{E}_{g_n}=\{\textbf{e}_{u_1},\textbf{e}_{u_2},...,\textbf{e}_{u_K}\}$, which will be aggregated by the LLM $\Theta$, denoted as:
\begin{equation}\label{eq:pa}
  \textbf{e}'_{g_n} = \Theta(\textbf{E}_{g_n}).
\end{equation}
Please note that $\textbf{E}_{g_n}$ is the input for the first transformer decoder of $\Theta$ without word embedding and positional encoding. This is because $\textbf{E}_{g_n}$ is a set of user representations instead of words, and the order of users is not important for preference aggregation. The output of the final transformer decoder will be utilized as the representation of the aggregated preference. Then, the enhanced group representation is obtained by:
\begin{equation}
  \textbf{e}_{g_n} = \textbf{e}_{g_n} + \alpha \textbf{e}'_{g_n},
\end{equation}
where $\alpha$ is a hyperparameter that controls the weight of the aggregated member preference. 

\subsubsection{Fine-tuning with QLoRA for Representation Refinement}
In section \ref{sec:lse}, we have trained an adapter $\Theta^{'}_s$ for generating check-in sequence representation. Obviously, this adapter cannot be applied directly to aggregate individual preferences. To distinguish between these two tasks and avoid compromising the performance of the sequencing adapter $\Theta^{'}_s$, we introduce a new adapter $\Theta^{'}_a$ with the same architecture as $\Theta^{'}_s$. On this basis, Equation \ref{eq:pa} is modified as:
\begin{equation}
  \textbf{e}'_{g_n} = \Theta_a(\textbf{E}_{g_n}),
\end{equation}
where $\Theta_a$ is the modified large language model by the aggregation adapter $\Theta^{'}_a$. During the fine-tuning process, while freezing all other weights, $\Theta^{'}_a$ is optimized by the POI prediction objective shown in Equation \ref{eq:lpoi}, only with the group check-in sequences and the corresponding group member check-in sequences. Here, we denote the set of all group check-in sequences and their members' check-in sequences as $\mathcal{D}_{group}$.

\subsection{Addressing Data Sparsity with Self-supervised Leaning}\label{sec:ssl}

As mentioned above, the LLM will be fine-tuned for check-in sequence representation generation and individual preference aggregation. However, the sparsity of user-POI interactions has been a long-standing issue in POI recommendation, which deteriorates further in group settings. To this end, we introduce a self-supervised signal to enhance representations for check-in sequences with the corresponding purposes, such as business trips and tourism, as the travel purposes provide a strong signal to predict the next POI a group/user will visit. Unfortunately, check-in data typically lacks explicit labels indicating the purpose of each check-in sequence. To overcome this limitation, we leverage the powerful capabilities of the LLM to classify each check-in sequence. The LLM analyzes the spatio-temporal and contextual information within the sequence and assigns a category that reflects its underlying purpose. To achieve this, we design the corresponding prompt which is shown as follows:
\small
\setlength{\fboxsep}{5pt} 
\setlength{\fboxrule}{0pt} 
\noindent 
\begin{promptbox}
\textit{Given a sequence of check-in activities, your task is to classify into one of the following
categories: \\\textcolor{purple}{\{Business Trip, Work Commute, Tourism, Short Vacation, Cultural \& Entertainment,
Shopping \& Dining, Social Events, Fitness \& Wellness, Daily Commute, Medical \& Health
Visits, Education \& Training\}.} \\\\
The format of a check-in activity is:
(POIID, Name, Category, longitude, latitude, temporal difference from previous check-in, spatial
difference from previous check-in)\\
The check-in sequence is: \textcolor{teal}{<check-in sequence>}\\
The format of your response: This sequence is classified as \textcolor{olive}{[Category]}.}
\end{promptbox}

As shown in the above prompt, we have $11$ purpose labels: 
\{Business Trip, Work Commute, Tourism, Short Vacation, Cultural \& Entertainment, Shopping \& Dining, Social Events, Fitness \& Wellness, Daily Commute, Medical \& Health Visits, Education \& Training\}. These labels are designed to reflect the most common and distinct purposes for short-term trips or sequences of check-ins. Each category is tied to a specific motivation or context that drives user mobility, ranging from daily routines to more occasional events. Together, they comprehensively capture the various reasons behind short-duration check-ins, ensuring that each sequence can be classified meaningfully and relevantly for POI prediction. Beyond that, those purpose labels are only valid for short sequences, since a purposeful trip won't last long. Therefore, we divide the whole sequence into subsequences if the time difference between any two consecutive check-in activities is larger than five days. Here, we denote the set of all short sequences as $\mathcal{D}_{ssl}$. To verify the accuracy of sequence purposes generated by the LLM, we manually label a portion ($20\%$) of short sequences as the ground truths. The LLM achieves the desirable accuracy of 92\%.

To endow our model with the ability for purpose prediction, we propose a new purpose embedding matrix $\textbf{E}_{pur} \in \mathbb{R}^{d\times L}$, where $L=11$ refers to the number of purpose labels. Consequently, given a check-in sequence representation $\textbf{e}_{\mathcal{X}}$, the probability of purpose labels is calculated as:
\begin{equation}
  \textbf{b} = \textbf{E}_{pur} \textbf{e}^{T}_{\mathcal{X}}.
\end{equation}
Then, the purpose prediction objective is defined as:
\begin{equation}
  \mathcal{L}_{pur}=\mathcal{L}_{CE}(y_{p}^{'},y_p),
\end{equation}
where $\mathcal{L}_{CE}$ is the cross-entropy loss, $y_{p}^{'}$ is the predicted purpose by our model, and $y_p$ is the purpose label generated by a standard but more powerful LLM (e.g., GPT-4). This loss will be used to pre-train the sequencing adapter $\Theta^{'}_s$ and the POI embedding matrix $\textbf{E}_{pur}$, and thus, the sparse group check-in data can leverage the relatively rich individual check-in data to obtain richer representations. The additional semantic information provided by the LLM enables the model to infer latent intents behind check-in sequences, leading to more precise and intent-aware POI recommendations.

\subsection{Optimization}\label{sec:opt}
\begin{algorithm}
  \caption{The optimization of LLMGPR.}
  \label{alg:opt}
\begin{algorithmic}[1]
    \STATEx /*Pretrain weights for sequence representation generation */
    \STATE Obtain the large language model $\Theta$;
    \STATE Initialize $\Theta^{'}_s$, $\Theta^{'}_a$, $\textbf{E}_{poi}$, $\textbf{E}_{pur}$;
    \STATE Obtain $\Theta_s$ by combining $\Theta$ and $\Theta^{'}_s$;  
    \STATE Obtain $\Theta_a$ by combining $\Theta$ and $\Theta^{'}_a$;
    \REPEAT
        \FOR{$(\mathcal{X},y_p)\in\mathcal{D}_{ssl}$}
            \STATE $y'_{p} \leftarrow \Theta_s(\mathcal{X},\textbf{E}_{poi},\textbf{E}_{pur})$;
            \STATE Take a gradient step w.r.t $\mathcal{L}_{pur}(y'_p,y_p)$ to update $\Theta^{'}_s$, $\textbf{E}_{poi}$, $\textbf{E}_{pur}$;
        \ENDFOR
    \UNTIL{convergence}
    \STATEx /*Train weights for sequence representation generation */
    \REPEAT
        \FOR{$(\mathcal{X},y)\in\mathcal{D}$}
            \STATE $y' \leftarrow \Theta_s(\mathcal{X},\textbf{E}_{poi})$;
            \STATE Take a gradient step w.r.t $\mathcal{L}_{poi}(y',y)$ to update $\Theta^{'}_s$, $\textbf{E}_{poi}$;
        \ENDFOR
    \UNTIL{convergence}
    \STATEx /*Train the aggregation adapter*/
    \REPEAT
        \FOR{$(\mathcal{X}_{g_n},\mathcal{K}_n,y)\in\mathcal{D}_{group}$}
            \STATE $\textbf{e}_{g_n}=\Theta_s(\mathcal{X}_{g_n},\textbf{E}_{poi})$
            \FOR{$\mathcal{X}_{u_k}\in\mathcal{X}(\mathcal{K}_n)$}
                \STATE $\textbf{e}_{u_k}=\Theta_s(\mathcal{X}_{u_k},\textbf{E}_{poi})$;
            \ENDFOR
            \STATE $\textbf{E}_{g_n}=\{e_{u_1},e_{u_2},...,e_{u_K}\}$;
            \STATE $\textbf{e}'_{g_n}=\Theta_a(\textbf{E}_{g_n})$;
            \STATE $\textbf{e}_{g_n} = \textbf{e}_{g_n} + \alpha \textbf{e}'_{g_n}$;
            \STATE $y' \leftarrow \textbf{E}_{poi}(\textbf{e}_{g_n})$;
            \STATE Take a gradient step w.r.t $\mathcal{L}_{poi}(y',y)$ to update $\Theta^{'}_a$;
        \ENDFOR
    \UNTIL{convergence}
\end{algorithmic}
\end{algorithm}

The optimization workflow of LLMGPR is presented in Algorithm \ref{alg:opt}. At the very beginning (line 1), we regard Llama3-8b as the base large language model $\Theta$, which can be replaced by any other large language model if sufficient computational resources are available. After that, we initialize all trainable weights including the check-in sequencing adapter $\Theta^{'}_s$, the aggregation adapter $\Theta^{'}_a$, the POI embedding matrix $\textbf{E}_{poi}$, and the purpose embedding matrix $\textbf{E}_{pur}$ (line 2). Then, we obtain the sequence-specific LLM $\Theta_s$ and the aggregation-specific LLM $\Theta_a$ (line 3-4). On this basis, we first pre-train the sequencing adapter $\Theta^{'}_s$ and the POI embedding matrix $\textbf{E}_{poi}$ based on the purpose prediction loss $\mathcal{L}_{pur}(y'_p,y_p)$ using all short sequences and their corresponding purpose labels (lines 5-10). The sequencing adapter $\Theta^{'}_s$ and the POI embedding matrix $\textbf{E}_{poi}$ are further optimized by the POI prediction loss $\mathcal{L}_{poi}(y',y)$ (lines 11-16). To address the group check-in data sparsity and obtain a better task-specific LLM for learning the representation of check-in sequences, all check-in sequences are utilized including group and all individual check-in sequences.

Subsequently, the aggregation adapter $\Theta^{'}_a$ is trained iteratively with group check-in sequences and corresponding individual check-in sequences of each group member (lines 17-29). In detail, for each group check-in sequence, the group representation is originally obtained by applying the sequence-specific LLM $\Theta_s$ to the group check-in sequence (line 19). After that, individual representations for group members are obtained by applying $\Theta_s$ to their respective check-in sequences (lines 20-23). These representations are then fed into the aggregation-specific LLM $\Theta_a$ (line 24), while the aggregated representation is utilized to enhance the group representation (line 25) and generate the recommendation results (line 26). Finally, the aggregation adapter is updated using the POI prediction loss $\mathcal{L}_{poi}(y',y)$ (line 27).

\section{Experiments}

In this section, we perform comprehensive experiments with three real-world datasets to evaluate the effectiveness of the proposed LLMGPR. Specifically, our investigation seeks to address the following research questions:

\noindent{\textbf{RQ1}}: How effective is LLMGPR for group POI recommendations?

\noindent{\textbf{RQ2}}: How does LLMGPR perform in learning check-in sequence representation compared to advanced POI recommender systems?

\noindent{\textbf{RQ3}}: Is each key component in LLMGPR effective?

\noindent{\textbf{RQ4}}: Does LLMGPR work well for cold-start recommendations?

\noindent{\textbf{RQ5}}: What is the influence of LLMGPR's key hyperparameters?

\begin{table}[htbp]
\footnotesize
  \centering
  \caption{Dataset statistics.}
  \label{table:data}
  \resizebox{\linewidth}{!}{
    \begin{tabular}{lrrr}
          \hline
          & \multicolumn{1}{r}{Foursquare} & \multicolumn{1}{r}{Weeplace} & \multicolumn{1}{r}{Gowalla}\\
    \hline
    \#users & 7,507    & 4,560 & 31,751\\
    \hline
    \#groups & 1,715     & 923 & 2186\\
    \hline
    \#POIs & 80,962     & 44,194 & 81,123\\
    \hline
    \#categories & 436     & 625 & 537\\
    \hline
    \#user check-ins & 1,214,631     & 623,654 & 862,502\\
    \hline
    \#group check-ins  & 12,594     & 11,974 & 7,738\\
    \hline
    \#check-ins per user & 162.80   & 136.77 & 27.16\\
    \hline
    \#check-ins per group & 7.34   & 12.97 & 3.54\\
    \hline
    \#users per group & 3.72   & 4.37 & 2.95\\
    \hline
    \end{tabular}%
    }
\end{table}%

\subsection{Datasets and Evaluation Protocols}\label{sec:dataset}

We adopt three real-world datasets including Foursquare \cite{2020Will}, Weeplace \cite{long2023model}, and Gowalla \cite{2013Personalized} to evaluate the proposed LLMGPR, where all datasets include user's check-in histories in the cities of New York, Los Angeles, and Chicago. Following \cite{Li2018NextPR,long2024physical}, users and POIs with less than 10 interactions are removed. Since all datasets do not originally contain group information, we follow the widely adopted procedure \cite{chen2022thinking} to construct group interactions by finding overlaps in check-in times and social relations. That is, we assume if a set of users who are connected on the social network visit the same venue at the same time, then they are regarded as members of a group, and the corresponding activities are group activities. The use of explicit social connections and spatiotemporal tags ensures the quality of discovered user groups. Table \ref{table:data} summarizes the statistics of the three datasets. 

For evaluation, we adopt the leave-one-out protocol which is widely used in previous works \cite{zheng2016keyword,zhang2023comprehensive}. That is, for each of the check-in sequences, the last check-in POI is for testing, the second last POI is for validation, and all others are for training. In addition, the maximum sequence length is set to 200. For each ground truth check-in POI, unlike the E-commerce setting that ranks all products~\cite{2020On,yin2024device}, we only pair it with 500 unvisited and nearest POIs within the same region as the candidates for ranking. The rationale is, different from e-commerce products \cite{2020On,qu2021imgagn}, in the scenario of POI recommendations that are location-sensitive, users seldom travel between two POIs consecutively that are far away from each other \cite{long2023decentralized,li2021discovering}. On this basis, we leverage two ranking metrics, namely Hit Ratio at Rank $k$ (HR@$k$) and Normalized Discounted Cumulative Gain at Rank $k$ (NDCG@$k$) \cite{2007CoFiRank} where HR@$k$ only measures the times that the ground truth is present on the top-$k$ list, while NDCG@$k$ cares whether the ground truth can be ranked as highly as possible.

For hyperparameters, the dimension size of POI embeddings is set to 4096, which is the same as that of original token embeddings in Llama3-8b. With respect to QLoRA, the quantization bit width $b$ is set to $4$ and the rank $r$ is set to $16$. Besides, we set $\alpha$ to 0.7, the learning rate to $0.0002$, the dropout to $0.2$, the batch size to $16$, and the maximum training epoch to $5$. The impacts of $r$ and $\alpha$ will be further discussed in Section \ref{sec:rq5}.

\begin{table*}[t]
\footnotesize
  \centering
  \caption{Recommendation performance comparison with POI recommenders for groups.}
  \label{table:grouprec}
  \resizebox{\linewidth}{!}{
    \begin{tabular}{|l|cccc|cccc|cccc|}
      \hline
          & \multicolumn{4}{c|}{Foursquare}  & \multicolumn{4}{c|}{Weeplace} & \multicolumn{4}{c|}{Gowalla} \\
          \cline{2-13}
          & \multicolumn{1}{c}{HR@5} & \multicolumn{1}{c}{NDCG@5} & \multicolumn{1}{c}{HR@10} & \multicolumn{1}{c|}{NDCG@10} & \multicolumn{1}{c}{HR@5} & \multicolumn{1}{c}{NDCG@5} & \multicolumn{1}{c}{HR@10} & \multicolumn{1}{c|}{NDCG@10} & \multicolumn{1}{c}{HR@5} & \multicolumn{1}{c}{NDCG@5} & \multicolumn{1}{c}{HR@10} & \multicolumn{1}{c|}{NDCG@10} \\
          \hline
          MF-AVG    & 0.2530 & 0.1457 & 0.3577 & 0.2114 & 0.2578 & 0.1476 & 0.3752 & 0.2279 & 0.2050 & 0.1238 & 0.3205 & 0.1875 \\
          AGREE     & 0.2731 & 0.1573 & 0.3869 & 0.2303 & 0.2612 & 0.1552 & 0.4201 & 0.2459 & 0.2123 & 0.1306 & 0.3486 & 0.2038 \\
          GroupIM   & 0.2863 & 0.1639 & 0.3872 & 0.2420 & 0.2848 & 0.1652 & 0.4431 & 0.2498 & 0.2230 & 0.1314 & 0.3595 & 0.2009 \\
          CubeRec   & 0.2919 & 0.1638 & 0.4016 & 0.2454 & 0.2826 & 0.1682 & 0.4435 & 0.2582 & 0.2288 & 0.1402 & 0.3656 & 0.2061 \\
          HHGR      & 0.3057 & 0.1705 & 0.4247 & 0.2584 & 0.3029 & 0.1789 & 0.4409 & 0.2595 & 0.2425 & 0.1495 & 0.3718 & 0.2178 \\
          MICL      & 0.3068 & 0.1711 & 0.4291 & 0.2571 & 0.3043 & 0.1786 & 0.4672 & 0.2676 & 0.2565 & 0.1477 & 0.3808 & 0.2299 \\
          LLMGPR    & \textbf{0.3182} & \textbf{0.1817} & \textbf{0.4767} & \textbf{0.2760} & \textbf{0.3201} & \textbf{0.1856} & \textbf{0.4799} & \textbf{0.2829} & \textbf{0.2739} & \textbf{0.1619} & \textbf{0.4203} & \textbf{0.2512} \\
          \hline
    \end{tabular}%
    }
\end{table*}%

\begin{table*}[t]
\footnotesize
  \centering
  \caption{Recommendation performance comparison with POI recommenders for individuals.}
  \label{table:personal}
  \resizebox{\linewidth}{!}{
    \begin{tabular}{|l|cccc|cccc|cccc|}
      \hline
          & \multicolumn{4}{c|}{Foursquare}  & \multicolumn{4}{c|}{Weeplace} & \multicolumn{4}{c|}{Gowalla} \\
          \cline{2-13}
          & HR@5 & NDCG@5 & HR@10 & NDCG@10 & HR@5 & NDCG@5 & HR@10 & NDCG@10 & HR@5 & NDCG@5 & HR@10 & NDCG@10 \\
          \hline
          GeoMF    & 0.1029 & 0.0590 & 0.1563 & 0.1075 & 0.0845 & 0.0599 & 0.1344 & 0.0937 & 0.0706 & 0.0634 & 0.1238 & 0.0955 \\
          STAN     & 0.3112 & 0.1770 & 0.4619 & 0.2759 & 0.2930 & 0.1734 & 0.4155 & 0.2547 & 0.2560 & 0.1520 & 0.3601 & 0.2164 \\
          DRAN     & 0.3020 & 0.1744 & 0.4769 & 0.2941 & 0.3064 & 0.1726 & 0.4220 & 0.2625 & 0.2579 & 0.1564 & 0.3751 & 0.2335 \\
          Diff-POI & 0.3119 & 0.1885 & 0.4711 & 0.2787 & 0.3034 & 0.1760 & 0.4681 & 0.2822 & 0.2585 & 0.1543 & 0.4050 & 0.2415 \\
          META ID  & 0.2947 & 0.1721 & 0.4696 & 0.2780 & 0.3064 & 0.1723 & 0.4129 & 0.2615 & 0.2557 & 0.1545 & 0.3728 & 0.2381 \\
          LLMMove  & 0.3127 & 0.1799 & 0.4693 & 0.2805 & 0.2958 & 0.1770 & 0.4627 & 0.2657 & 0.2540 & 0.1572 & 0.4044 & 0.2428 \\
          LLM4POI  & 0.3173 & 0.1948 & 0.4776 & 0.2849 & 0.3023 & 0.1874 & 0.4703 & 0.2891 & 0.2699 & 0.1628 & 0.4246 & 0.2492 \\
          LLMGPR   & \textbf{0.3233} & \textbf{0.2040} & \textbf{0.4872} & \textbf{0.2995} & \textbf{0.3196} & \textbf{0.1963} & \textbf{0.4867} & \textbf{0.2920} & \textbf{0.2853} & \textbf{0.1958} & \textbf{0.4483} & \textbf{0.2586} \\
          \hline
        \end{tabular}%
        }
\end{table*}

\subsection{Baselines}
To evaluate the effectiveness of LLMGPR for group POI recommendations, we first compare it with existing group recommenders. For a fair comparison, we regard check-in sequence representations and POI representations generated by the well-trained LLM as initial representations of groups, users, and POIs for the above baselines. To validate the effectiveness of LLMGPR in learning check-in sequence representations, we also compare it with SOTA POI recommenders for individuals. All baselines are summarized below:

\noindent\textbf{Group POI Recommenders:}
\begin{itemize}[leftmargin=*]
  \item \textbf{MF-AVG} \cite{vinh2019interact}: It averages group-wise user representations learned via user-item matrix factorization to form group embeddings.
  
  \item \textbf{AGREE} \cite{cao2018attentive}: It incorporates an attention mechanism to deduce group preferences from individual member preferences and aggregates them into a unified group preference vector.
  
  \item \textbf{GroupIM} \cite{sankar2020groupim}: It Utilizes expressive hypercubes in vector space with a new distance metric and an intersection-based self-supervision paradigm to facilitate group-item pairwise ranking.

  \item \textbf{CubeRec} \cite{chen2022thinking}: It utilizes expressive hypercubes in vector space with a new distance metric and an intersection-based self-supervision paradigm to facilitate group-item pairwise ranking.

  \item \textbf{HHGR} \cite{zhang2021double}: It employs a hypergraph neural network to encapsulate complex high-order relationships among users, groups, and items, with a double-scale self-supervised learning objective.

  \item \textbf{MICL} \cite{li2024multi}: It employs the group-view adaptive graph transformer, member-view hypergraph aggregation network, and item-view tripartite graph augmentation to deeply explore potential interactive compromises.
\end{itemize} 

\noindent\textbf{Normal POI Recommenders:}
\begin{itemize}[leftmargin=*]

  \item \textbf{MF} \cite{2014GeoMF}: It utilizes user-item matrix factorization for POI recommendations.
  
  \item \textbf{STAN} \cite{2021STAN}: It learns explicit spatiotemporal correlations of check-in trajectories via a bi-attention approach.

  \item \textbf{DRAN} \cite{10.1145/3477495.3532012}: It leverages a disentangled representation-enhanced attention network for next POI recommendation using GNN-based methods.

  \item \textbf{Diff-POI} \cite{Qin_2023}: It uses a diffusion-based model to sample from the posterior distribution that reflects user geographical preferences.
  
  \item \textbf{META ID} \cite{huang2024improving}: It introduces ID representations in LLM-based recommender systems, and uses out-of-vocabulary tokens to characterize users/items, enhancing the memorization and diversity of their representations.

  \item \textbf{LLMMove} \cite{feng2024move}: It applies zero-shot generalization of the LLM for next POI recommendations.

  \item \textbf{LLM4POI} \cite{li2024large}: It fine-tunes the LLM on standard-sized datasets to exploit commonsense knowledge for the next POI recommendations.
  
\end{itemize}

\subsection{Group Recommendation Effectiveness (RQ1)}\label{sec:rq1}

The performance comparison across all group recommenders, summarized in Table \ref{table:grouprec}, provides several key insights. Firstly, AGREE consistently outperforms MF-AVG across all datasets, benefiting from its attention mechanism that dynamically assigns weights for preference aggregation, thereby capturing interactions and relationships among group members more effectively. However, its reliance on point embeddings limits flexibility and adaptability in group representation. To address this, GroupIM and CubeRec introduce hypercube-based group representations, replacing point embeddings with range embeddings, which improve recommendation accuracy. Additionally, HHGR and MICL leverage graph neural networks to model relationships between users, groups, and POIs, leading to more precise group representations and enhanced recommendation performance.

Despite these advances, the proposed LLMGPR model demonstrates superior accuracy across all datasets, outperforming all baselines. Specifically, compared to the best-performing baseline, MICL, LLMGPR achieves average improvements of 7.09\% on Foursquare, 4.39\% on Weeplace, and 9.01\% on Gowalla. In addition, Gowalla, characterized by its extremely sparse group-POI interactions, presents significant challenges for recommendation models. Nonetheless, LLMGPR effectively leverages enriched contextual information and augmented supervision signals to learn high-quality group representations, achieving remarkable accuracy even on this challenging dataset. This highlights the robustness and effectiveness of LLMGPR, particularly in addressing data sparsity issues that hinder other methods.


\begin{table*}[t]
\footnotesize
  \centering
  \caption{Ablation study for group POI recommendation}
  \label{table:ablation}
  \resizebox{\linewidth}{!}{
    \begin{tabular}{|l|cccc|cccc|cccc|}
      \hline
          & \multicolumn{4}{c|}{Foursquare} & \multicolumn{4}{c|}{Weeplace} & \multicolumn{4}{c|}{Gowalla} \\
          \cline{2-13}
          & HR@5 & NDCG@5 & HR@10 & NDCG@10 & HR@5 & NDCG@5 & HR@10 & NDCG@10 & HR@5 & NDCG@5 & HR@10 & NDCG@10 \\
          \hline
          LLMGPR      & \textbf{0.3182} & \textbf{0.1817} & \textbf{0.4767} & \textbf{0.2760} & \textbf{0.3201} & \textbf{0.1856} & \textbf{0.4799} & \textbf{0.2829} & \textbf{0.2739} & \textbf{0.1619} & \textbf{0.4203} & \textbf{0.2512} \\
          LLMGPR-FT   & 0.2672 & 0.1619 & 0.3978 & 0.2283 & 0.2892 & 0.1625 & 0.4301 & 0.2321 & 0.2239 & 0.1436 & 0.3381 & 0.2006 \\
          LLMGPR-M    & 0.2786 & 0.1623 & 0.4176 & 0.2459 & 0.2814 & 0.1642 & 0.4362 & 0.2569 & 0.2301 & 0.1408 & 0.3526 & 0.2203 \\
          LLMGPR-ER   & 0.2819 & 0.1688 & 0.4227 & 0.2405 & 0.2832 & 0.1659 & 0.4306 & 0.2573 & 0.2381 & 0.1446 & 0.3673 & 0.2116 \\
          LLMGPR-SSL  & 0.2989 & 0.1786 & 0.4541 & 0.2665 & 0.3027 & 0.1790 & 0.4502 & 0.2753 & 0.2517 & 0.1528 & 0.3739 & 0.2370 \\
          \hline
    \end{tabular}%
    }
\end{table*}%

\begin{table*}[htbp]
  \footnotesize
  \centering
  \caption{Cold-start recommendation performance.}
  \label{table:rq4}
  \resizebox{\linewidth}{!}{
    \begin{tabular}{|l|cccc|cccc|cccc|}
          \hline
          Model & \multicolumn{4}{c}{Foursquare} & \multicolumn{4}{c}{Weeplace} & \multicolumn{4}{c|}{Gowalla} \\
          \cline{2-13}
           & HR@5 & NDCG@5 & HR@10 & NDCG@10 & HR@5 & NDCG@5 & HR@10 & NDCG@10 & HR@5 & NDCG@5 & HR@10 & NDCG@10 \\
    \hline
    MF-AVG & 0.2078 & 0.1223 & 0.2983 & 0.1793 & 0.2148 & 0.1254 & 0.3134 & 0.1966 & 0.1681 & 0.1044 & 0.2730 & 0.1630 \\
    AGREE  & 0.2279 & 0.1362 & 0.3250 & 0.1949 & 0.2206 & 0.1312 & 0.3449 & 0.2120 & 0.1815 & 0.1111 & 0.2908 & 0.1761 \\
    GroupIM & 0.2396 & 0.1357 & 0.3301 & 0.2030 & 0.2468 & 0.1411 & 0.3664 & 0.2113 & 0.1883 & 0.1115 & 0.2950 & 0.1747 \\
    CubeRec & 0.2450 & 0.1393 & 0.3470 & 0.2033 & 0.2355 & 0.1443 & 0.3758 & 0.2218 & 0.1878 & 0.1155 & 0.3068 & 0.1709 \\
    HHGR   & 0.2618 & 0.1455 & 0.3572 & 0.2244 & 0.2605 & 0.1480 & 0.3680 & 0.2181 & 0.2103 & 0.1248 & 0.3144 & 0.1838 \\
    MICL   & 0.2532 & 0.1438 & 0.3658 & 0.2156 & 0.2573 & 0.1531 & 0.4064 & 0.2258 & 0.2152 & 0.1220 & 0.3208 & 0.1857 \\
    LLMGPR & \textbf{0.3029} & \textbf{0.1780} & \textbf{0.4630} & \textbf{0.2659} & \textbf{0.3070} & \textbf{0.1751} & \textbf{0.4686} & \textbf{0.2696} & \textbf{0.2577} & \textbf{0.1515} & \textbf{0.4051} & \textbf{0.2395} \\
    \hline
    \end{tabular}%
    }
\end{table*}%

\subsection{Effectiveness of Learning Check-in Sequence Representations (RQ2)}\label{sec:rq2}

To further study the effectiveness of the proposed LLMGPR in learning check-in sequence representation, we also compare LLMGPR with SOTA POI recommenders for individuals, where the results are summarized in Table \ref{table:personal}. Our proposed LLMGPR consistently and significantly beats the baselines on all datasets. As the basic POI recommendation approach without considering the sequential effect of user check-in activities, GeoMF undoubtedly has the worst performance on all datasets. STAN outperforms GeoMF by a significant margin since it can leverage spatiotemporal correlations in both consecutive and non-consecutive check-in activities with the attention mechanism. Advancing further, DRAN melds a Graph Neural Network (GNN) with an attention mechanism, leading to more refined POI embeddings and thus outperforming STAN in terms of accuracy. Most notably, Diff-POI, employing the robust generality of the diffusion model, sets a new benchmark for state-of-the-art accuracy in this domain. 

Recent works focus on integrating the LLM for recommendations, due to its ability to leverage the semantic information of users and items. META ID  uses out-of-vocabulary tokens to characterize users and items for recommendations. As it is not designed for POI recommendations, it cannot integrate and exploit the spatial and temporal information of check-ins, even performing worse than Diff-POI. In addition, LLMMove applies the zero-shot generation ability of the LLM for the next POI recommendations. Without fine-tuning for the specific recommendation task, its performance is limited. Instead, LLM4POI fine-tunes the LLM on the three datasets for the POI recommendation task, and thus, it achieves better performance. Notably, our proposed LLMGPR significantly outperforms the best baseline LLM4POI, showing its superiority in learning check-in sequence representations. This is attributed to the well-trained sequencing adapter and the introduction of a novel self-supervised learning task.


\subsection{Ablation Study (RQ3)}
In this section, we perform the ablation analysis to evaluate the performance gain from the core components of LLMGPR in terms of group POI Recommendation. Table \ref{table:ablation} summarizes the recommendation outcomes for different degraded versions of LLMGPR. Then, we introduce all variants and discuss the effectiveness of corresponding model components.

\textbf{Using the LLM directly without fine-tuning (LLMGPR-FT).} In LLMGPR, we modify and fine-tune the LLM for group POI recommendations. To evaluate the usefulness of those modifications, we apply the original LLM directly for group POI recommendations, where both group-level and member-level check-in sequences are utilized. This variant model has experienced noticeable performance drops, especially on the Gowalla dataset with the higher sparsity of group-item interactions. Hence, including modification and fine-tuning of the LLM is important for LLMGPR to learn representative group representations in the scenario of data sparsity.

\textbf{Using the fine-tuned but not modified LLM (LLMGPR-M).} This variant differs from LLMGPR-FT as it is fine-tuned using the QLoRA strategy. It can also be seen as directly applying LLM4POI \cite{li2024large} for group POI recommendations. As anticipated, the recommendation accuracy drops significantly, underscoring the effectiveness of the additional POI embedding matrix and the innovative approach to constructing group representations.

\textbf{Removing the enhancement of group representation with aggregated member preferences (LLMGPR-ER).}
A crucial component of LLMGPR is enhancing group representation by aggregated member preference as described in Section \ref{sec:er}. To test its effectiveness, we remove it and only utilize the group check sequence for final recommendations. As a result, there is a significant decrease in recommendation accuracy for all datasets. This reflects that, even with the well-trained LLM, the sparse group-item interactions significantly harm the accuracy, and our proposed aggregation strategy can effectively address this issue.

\textbf{Removing self-supervised learning (LLMGPR-SSL).}
This variant disables the self-supervised loss when pretraining the sequencing adapter $\Theta^{'}_s$ and the POI embedding matrix $\textbf{E}_{poi}$, as shown in Section \ref{sec:ssl}. As LLMGPR can no longer leverage the purposes of short sequences as a supervision signal to counteract the group-level data sparsity, it suffers from inferior recommendation accuracy on all datasets. Therefore, the novel self-supervised learning scheme showcases its strong contribution to enriching group-level information.

\subsection{Effect of Cold-start Recommendations (RQ4)}

To evaluate the effectiveness of LLMGPR in cold-start scenarios, we conducted experiments using 200 randomly selected group-level check-in sequences from each dataset. These sequences, with fewer than 10 check-ins each, were excluded from the training process to simulate cold-start conditions. The results, presented in Table \ref{table:rq4}, demonstrate that LLMGPR consistently achieves superior performance compared to baseline models across all datasets. Furthermore, compared to its performance on non-cold-start data (i.e., sequences included in training), LLMGPR exhibits the smallest drop in accuracy among all models. This indicates its robustness and adaptability in handling the challenges of cold-start recommendation. The relatively stable performance decline highlights LLMGPR's effectiveness in leveraging enriched contextual information and its capability to generalize well to unseen scenarios. Its consistent superiority across diverse datasets underscores its potential to address real-world cold-start issues effectively.



\subsection{Hyperparameter Sensitivity (RQ5)}\label{sec:rq5}
\begin{figure}
    \includegraphics[width=\linewidth]{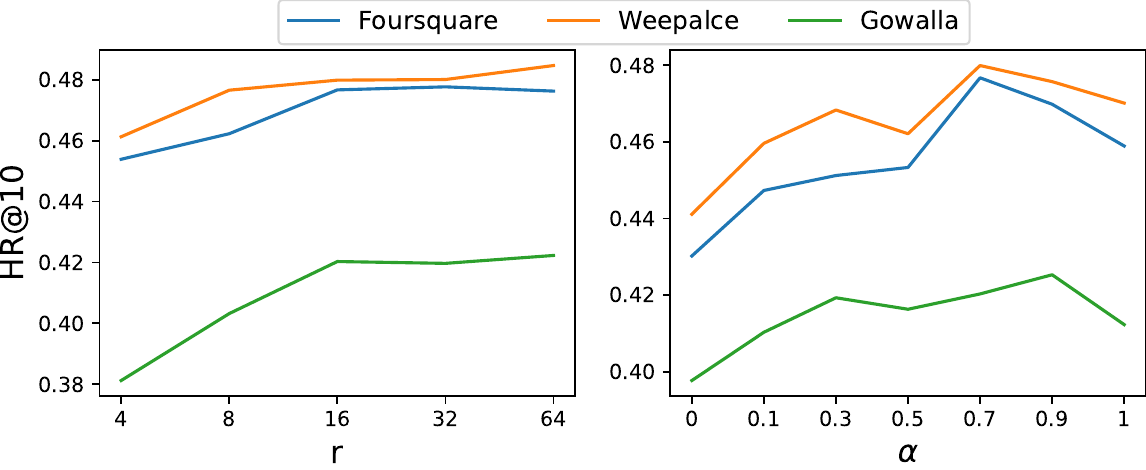}
	\caption{Hyperparameter sensitivity.}
    \label{fig:hs}
\end{figure}

In this section, we illustrate the effect of two hyperparameters on the group-level recommendation accuracy of LLMGPR including the rank $r$ which controls adapter sizes, and $\alpha$ that controls the injection level of the aggregated members' preference to the group representations. The results are shown in Figure \ref{fig:hs}.

\textbf{Impact of $r$.} Recommendation accuracy is recorded for $r\in\{4,8,16,32,64\}$. Usually, the recommendation accuracy benefits from higher $r$. However, the accuracy stabilizes once $r$ exceeds $16$, indicating that LLMGPR is capable of delivering high-performance recommendations without requiring additional resources to train larger adapters.

\textbf{Impact of $\alpha$.} We experiment on $\alpha\in\{0,0.1,0.3,0.5,0.7,0.9,1\}$. The lowest accuracy is obtained if the aggregated members' preference is not applied ($\alpha=0$), showing the significance of the strategy for enhancing group representations. At the start, the recommendation accuracy increases with the increase of $\alpha$. However, the accuracy will decline if knowledge of members' individual preferences has an excessive proportion ($\alpha>0.7$).

\section{Related work}
This section reviews recent literature on related areas including POI recommendations and group POI Recommendations.

\subsection{POI Recommendations}

To help people discover attractive places by analyzing user-POI interactions, early works on discovering attractive places through user-POI interactions primarily focused on matrix factorization \cite{2014GeoMF} and Markov chains \cite{2013Where,zheng2016keyword}. The advent of recurrent neural network (RNN) models marked a significant advancement, as they demonstrated the ability to capture spatiotemporal dynamics in POI sequences \cite{2018DeepMove, yuan2023interaction, li2021discovering}. Additionally, models utilizing attentive neural networks \cite{2021STAN, 10.1145/3477495.3531983, yin2015joint} adopted self-attention mechanisms to analyze the spatiotemporal context of sequential check-ins. Graph neural network (GNN)-based models \cite{li2021discovering,10.1145/3477495.3532012,gao2023graph,gao2023semantic} further advanced the field by integrating graph-augmented POI sequences, leveraging collaborative signals from semantically similar POIs to reveal sequential trends and outperform RNN-based approaches in accuracy. Diffusion-based methods \cite{Qin_2023, long2024diffusion} set a new standard to achieve cutting-edge accuracy. Recent research focuses on integrating Large Language Models (LLMs) for POI recommendations. LLMMove \cite{feng2024move} employs zero-shot generation for next POI recommendations, while LLM4POI \cite{li2024large} fine-tunes LLMs on related datasets to exploit commonsense knowledge. Compared to these approaches, LLMGPR offers substantial advantages in effectively learning sequence representations, enabling it to capture the intricate dynamics of user interactions. Additionally, it addresses group-level data sparsity by optimizing specialized adapters and enhancing the POI embedding matrix, resulting in more accurate and scalable recommendations.

\subsection{Group POI Recommendations}
Group POI recommenders typically involve two main phases: learning group representations from group-level check-in sequences and enhancing these representations using aggregated member representations derived from individual check-in sequences. Existing methods for preference aggregation range from simple averaging \cite{vinh2019interact}, which treats all members equally, to weighted averaging \cite{sojahrood2021poi}, which assigns weights based on external factors but struggles with determining appropriate values. More sophisticated approaches, such as attentive neural networks \cite{cao2018attentive, yin2019social}, hypercube-based embeddings \cite{sankar2020groupim,chen2022thinking}, and graph convolution networks \cite{zhang2021double,li2024multi}, aim to capture multifaceted preferences and complex relationships among users, groups, and POIs. However, these models often face challenges like overfitting and lack of scalability. In contrast, LLMGPR, a modified and well-trained LLM, is designed to capture complex patterns in user behavior and interactions. It dynamically adapts to individual preferences while offering enhanced performance and flexibility, making it particularly effective for group POI recommendations.

\subsection{LLM-based Recommendations}
Motivated by the notable achievements of LLMS, the integration of LLMs into recommendation systems has grown significantly. Early works \cite{feng2024move,hou2024large} primarily focused on designing prompts to utilize LLMS directly with zero-shot generalization. However, these approaches often struggled with the lack of specificity and domain adaptation, which led to the development of prompt tuning techniques \cite{zhai2023knowledge, wu2024personalized, huangcausal}. Prompt tuning involved adjusting the prompts to better suit the recommendation task, offering improved performance but still facing limitations in handling complex recommendation scenarios. This led to the exploration of fine-tuning LLMs \cite{lin2024data, bao2023tallrec}, where models were adapted to the recommendation tasks through task-specific training, significantly enhancing accuracy but lacking unique indicates of item IDs. Hence, the construction of IDs was introduced \cite{huang2024improving,yu2024ra}, creating embeddings or representations that could capture nuanced user preferences and item characteristics, thus offering a more efficient and scalable solution to recommendation problems. Building on these developments, LLMGPR introduces a novel framework that optimizes additional components, including POI embeddings, sequencing adapters, and aggregation adapters, to adapt LLMs for group-level recommendations. These enhancements aim to not only learn sequence representations of user interactions more effectively but also aggregate group members' preferences with greater accuracy, addressing the unique challenges of group recommendations and further pushing the boundaries of LLM-based recommendation systems.


\section{Conclusion}

In this paper, we introduced LLMGPR, a novel large language model-based framework specifically designed to tackle the complex challenges associated with group POI recommendation in Location-Based Social Networks. Our approach leverages the advanced capabilities of large language models to effectively learn nuanced sequence representations and dynamically aggregate the diverse preferences of group members. This allows LLMGPR to overcome critical limitations of existing methods, such as data sparsity, inadequate preference aggregation, and the inability to adapt to varying group dynamics. To enhance the performance of the framework, we introduced additional POI tokens and adapters that are meticulously designed to handle sequencing and aggregation tasks, enabling the model to generate dense, context-rich, and holistic representations of user preferences. Moreover, the incorporation of a self-supervised learning signal plays a crucial role in improving the framework's ability to generalize across diverse groups and contexts, ensuring robust performance in real-world applications. Our comprehensive experiments validate the effectiveness of LLMGPR, demonstrating its superiority over existing state-of-the-art methods in terms of recommendation accuracy and its ability to cater to the diverse and sometimes conflicting preferences of group members. These results highlight the transformative potential of LLMGPR in redefining group POI recommendation systems, offering a promising pathway to better satisfy user needs while addressing long-standing challenges in this domain. 



\begin{acks}
This work is supported by the Australian Research Council under the streams of Future Fellowship (Grant No. FT210100624),  the Discovery Project (Grant No. DP240101108), and the Linkage Projects (Grant No. LP230200892 and LP240200546).
\end{acks}

\bibliographystyle{ACM-Reference-Format}
\bibliography{sample-sigconf}

\end{document}